\documentclass[noshowpacs,prl,aps,superscriptaddress,floatfix,twocolumn]{revtex4}

\usepackage{lineno,hyperref,mathtools,doi}
\usepackage{amsmath}
\usepackage{amssymb,mathrsfs}
\usepackage{graphicx}
\usepackage{hyperref}
\usepackage{xcolor}

\DeclareMathAlphabet{\mathpzc}{OT1}{pzc}{m}{it}
\newcommand{\sss}{\scriptscriptstyle}

\begin{document}
\title{Departing from thermality of analogue Hawking radiation in a
  Bose-Einstein condensate}

\author{M. Isoard} \affiliation{Universit\'e Paris-Saclay, CNRS,
  LPTMS, 91405 Orsay, France}
\author{N. Pavloff} \affiliation{Universit\'e Paris-Saclay, CNRS,
  LPTMS, 91405 Orsay, France}

\begin{abstract}
  We study the quantum fluctuations in a one dimensional
  Bose-Einstein condensate realizing an analogous acoustic black
  hole. The taking into account of evanescent channels and of zero
  modes makes it possible to accurately reproduce recent experimental
  measurements of the density correlation function. We discuss the
  determination of Hawking temperature and show that in our model the
  analogous radiation presents some significant departure from
  thermality.
\end{abstract}

\maketitle

The Hawking effect \cite{Haw74} being of kinematic origin \cite{Vis98}
can be transposed to analogue systems, as first proposed by Unruh
\cite{Unr81}.  Among the various platforms which have been proposed
for observing induced or spontaneous analogous Hawking radiation and
related phenomena, the ones for which the experimental activity is
currently the most intense are surface water waves
\cite{Rou08,Wei11,Euv16,Car16,Euv18,Tor17,Goo19}, nonlinear light
\cite{Phi08,Bel10,Rub11,Ela12,Web14,Voc18,Dro19}, excitonic polaritons
\cite{Ngu15} and Bose-Einstein condensed atomic vapors
\cite{Lah10,Ste14,Ste16,Nov19}.

Because of their low temperature, of their intrinsic quantum nature,
and of the high experimental control achieved in these systems,
Bose-Einstein condensates (BECs) seem particularly suitable for
studying analogue Hawking effect. Steinhauer and colleagues have
undertaken several studies of quasi-unidimensional configurations
making it possible to realize analogue black hole horizons in BEC
systems, and made claims of observation of Hawking radiation
\cite{Ste14,Ste16,Nov19}. Their results have triggered the interest of
the community
\cite{Mic15,Mic16,Tet16,Nov16,Fin16,Wan17,Par17,Rob17,Fab18,Cou18,Gom19},
and generated a vivid debate \cite{Leo18,Ste18}. One of the goals of
the present Letter is to contribute to this debate, and to partially
close it, at least in what concerns density correlations around an
analogue black hole horizon. A definite theoretical answer can be
obtained thanks to a remark which had been overlooked in
previous works: one needs to develop the quasi-particle operator on
a complete basis set for properly describing the density
fluctuations. This is achieved in the first part of this letter, and
we apply this theoretical approach to the analysis of the
experimental results of Ref. \cite{Nov19}.

While in general relativity the thermality of the Hawking radiation is
constrained by the laws of black hole thermodynamics, no such general
principle is expected to hold for analogue systems \cite{Vis98}. It is
nonetheless commonly accepted that the spectrum of analogous Hawking
radiation only weakly departs from thermality\cite{Unr95,Cor96,Cor97},
and that all relevant features of an analogue system can be understood
on the basis of a hydrodynamical, long wave-length
description. However, the phenomenology of analogous systems provides
mechanisms supporting the impossibility of a perfectly thermal analogue
Hawking radiation \cite{Jac91}. In the second part of this Letter we
argue that in the BEC case we are considering, it is legitimate to
determine a Hawking temperature from the information encoded in the
density correlation function, but we show that some features of the
radiative process at hand significantly depart from thermality and we
propose a procedure for confirming our view.

We consider a one dimensional configuration in which the quantum field
$\hat{\Psi}(x,t)$ is solution of the Gross-Pitaevskii
equation
\begin{equation}\label{eq1}
i\hbar \partial_t \hat{\Psi}=-\frac{\hbar^2}{2m}\partial_x^2\hat{\Psi}
+\left[g \, \hat{n}+U(x)\right] \hat{\Psi}\; .
\end{equation}
In this equation $m$ is the mass of the atoms,
$\hat{n} = \hat{\Psi}^\dagger\hat{\Psi}$ and the term $g\, \hat{n}$
describes the effective repulsive atomic interaction ($g>0$).  We have
studied several external potentials $U(x)$ making it possible to
engineer a sonic horizon, but we only present here the results for a
step function: $U(x)=-U_0\Theta(x)$ with $U_0>0$. The reason for this
choice is twofold: (i) this potential has been realized experimentally
in Refs. \cite{Ste16,Nov19}, (ii) from the three configurations
analyzed in Ref. \cite{Fab18}, this is the one which leads to the
signal of quantum non-separability which is the largest and the most
resilient to temperature effects.

In the spirit of Bogoliubov's approach, we write the quantum field as
\begin{equation}
\hat\Psi(x,t)=\exp(-i\mu t/\hbar)\left[\Phi(x)+\hat\psi(x,t)\right],
\end{equation} 
where $\mu$ is the chemical potential. $\Phi(x)$ is a classical field
describing the stationary condensate and $\hat\psi(x,t)$ accounts for
small quantum fluctuations. Although such a separation is not strictly
valid in one dimension, it has been argued in Ref. \cite{Fab18} that it
constitutes a valid approximation over a large range of
one-dimensional densities. In the case we consider, $\Phi$ is a
solution of the classical Gross-Pitaevskii equation describing a sonic
horizon: the $x<0$ profile is half a dark soliton \cite{Leb03}, with
$\Phi(x\to-\infty)=\sqrt{n_u} \exp(i k_u x)$, where $n_u$ and
$V_u=m k_u/\hbar$ ($>0$) are the upstream asymptotic density and
velocity respectively. The downstream ($x>0)$ flow of the condensate
corresponds to a plane wave:
$\Phi(x>0) = \sqrt{n_d} \exp(i k_d x-i \pi/2)$. The asymptotic
upstream and downstream sound velocities are
$c_{(u,d)} = \sqrt{g n_{(u,d)}/m}$.  The analogous black hole
configuration corresponds to a flow which is asymptotically upstream
subsonic ($V_u<c_u$) and downstream supersonic
($\hbar k_d/m=V_d >c_d$).

\begin{figure}
\includegraphics[width=\linewidth]{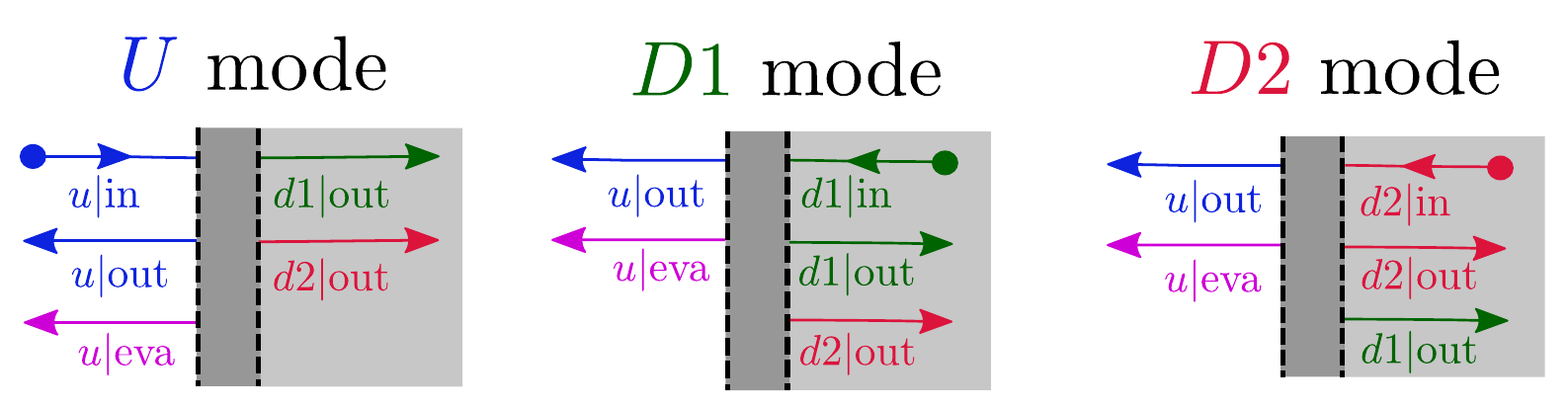}
\caption{Sketch of the different channels contributing to the incoming
  quantum modes $U$, $D1$ and $D2$. In each plot the background BEC
  propagates from left to right, the white region corresponds to the
  upstream subsonic flow, the gray one to the interior of the
  analogous black hole (downstream supersonic flow) and the region of
  the horizon is represented by the dark gray shaded interface. The
  Hawking channel and its partner are labeled $u|{\rm out}$ and
  $d2|{\rm out}$. The $d1|{\rm out}$ channel is a companion
  propagating away from the horizon, inside the analogous black hole
  region. Each mode ($U$, $D1$ and $D2$) is seeded by a ingoing
  channel ($u|{\rm in}$, $d1|{\rm in}$ and $d2|{\rm in}$) whose group
  velocity is directed towards the horizon.}
\label{fig1}
\end{figure}

We describe the quantum fluctuations on top of this classical field
within a linearized approach. The relevant modes are identified by
using the asymptotic ingoing (i.e. directed towards the acoustic
horizon) and outgoing channels, far from the horizon.  As discussed in
previous references \cite{Mac09,Rec09,Cou12,Lar12,Boi15} and recalled in
\cite{Supp}, the Bogoliubov dispersion relation supports a
decomposition of $\hat{\psi}$ onto three incoming modes which we
denote as $U$, $D1$ and $D2$.  For instance, the $U$ mode is seeded by
an upstream incoming wave which we denote as $u|{\rm in}$, which
propagates towards the horizon with a long wavelength group velocity
$V_u+c_u$. It is scattered onto two outgoing transmitted channels
(propagating in the analogue black hole away from the horizon) which we
denote as $d1|{\rm out}$ and $d2|{\rm out}$ with respective long
wavelength group velocities $V_d+c_d$ and $V_d-c_d$ (both positive)
and one outgoing reflected channel (propagating away from the horizon,
outside of the analogue black hole, with long wavelength group velocity
$V_u-c_u<0$). The corresponding three scattering coefficients are
denoted as $S_{d1,u}$, $S_{d2,u}$ and $S_{u,u}$. There is also an
upstream evanescent wave ($u|{\rm eva}$) which carries no current,
does not contribute to the $S$-matrix, but is important for fulfilling
the continuity relations at $x=0$. The situation is schematically
depicted in Fig. \ref{fig1}.

The frequency-dependent boson operators associated to the three
incoming modes $U$, $D1$ and $D2$ are denoted as $\hat{b}_{\sss U}$,
$\hat{b}_{\sss D1}$ and $\hat{b}_{\sss D2}$; they obey the commutation
relations
$[\hat{b}_{\sss L}(\omega) , \hat{b}_{\sss L'}^\dagger(\omega')] =
\delta_{\sss L,L'} \delta(\omega-\omega')$.
In addition, Bose-Einstein condensation is associated to a
spontaneously broken U(1) symmetry which implies the existence of
supplementary zero modes of the linearized version of \eqref{eq1}. As
discussed in Ref. \cite{Lew97}, one is lead to introduce two new
operators $\hat{\mathscr{P}}$ and $\hat{\mathscr{Q}}$ accounting for
the global phase degree of freedom, and the correct expansion of the
quantum fluctuation field reads
\begin{equation}\label{eq2}
\begin{split}
\hat{\psi}(x,t)= & - i \Phi(x) \hat{\mathscr{Q}} + 
i q(x) \hat{\mathscr{P}}
+ \int_0^\infty \frac{d\omega}{\sqrt{2\pi}}
\sum_{L\in\{U,D1\}} \\ 
& [u_{\sss L}(x,\omega) e^{-i\omega t} \, \hat{b}_{\sss L}(\omega)
+v_{\sss L}^*(x,\omega)e^{i\omega t} \, \hat{b}_{\sss L}^\dagger(\omega)]\\
& +\int_0^\Omega \frac{d\omega}{\sqrt{2\pi}}
[u_{\sss D2}(x,\omega) e^{-i\omega t} \, \hat{b}^\dagger_{\sss D2}(\omega)\\
& +v_{\sss D2}^*(x,\omega)e^{i\omega t}\, \hat{b}_{\sss D2}(\omega)].\\
\end{split}
\end{equation}
In this expression the $u_{\sss L}$'s and $v_{\sss L}$'s are the usual
Bogoliubov coefficients (their explicit form is given for instance in
Ref. \cite{Lar12}), and the quantization of the $D2$ mode is atypical,
as discussed in several previous references
\cite{Leo03,Mac09,Rec09}. The function $q(x)$ is one of the components
of the zero eigenmodes, see \cite{Supp}. Omitting the contribution of
the zero mode operators $\hat{\mathscr{P}}$ and $\hat{\mathscr{Q}}$
would correspond to using an incomplete basis set for the
expansion of the quantum fluctuations; in other words, their
contribution is essential for verifying the correct commutation
relation $[\hat{\psi}(x,t),\hat{\psi}^\dagger(y,t)]=\delta(x-y)$.  The
operator $\hat{\mathscr{Q}}$ is associated to the global phase of the
condensate. $\hat{\mathscr{P}}$ is the canonical conjugate operator
($[\hat{\mathscr{Q}},\hat{\mathscr{P}}]=i$) which typically appears in
the quadratic Hamiltonian $\hat{H}_{\rm quad}$ describing the dynamics
of the quantum fluctuations with a $\hat{\mathscr{P}}^2$ contribution,
while $\hat{\mathscr{Q}}$ does not \cite{Lew97,RS80,Bla86}. This means
that the degree of liberty associated to the broken symmetry has no
restoring force -- as expected on physical grounds -- and that the
zero excitation quantum state $|\rm{BH}\rangle$ describing the
analogous black hole configuration verifies
$\hat{\mathscr{P}}|\rm{BH}\rangle =0$ and
$\hat{b}_{\sss L}(\omega)|\rm{BH}\rangle=0$ for $L\in\{U,D1,D2\}$.

Once the appropriate expansion \eqref{eq2} has been performed, and the
correct quantum state $|\rm{BH}\rangle$ has been identified, one can
compute the density correlation function
\begin{equation}\label{eq3}
\begin{split}
G_2(x,y)=& \langle : \! \hat{n}(x,t)\hat{n}(y,t)\! : \rangle -
\langle \hat{n}(x,t)\rangle\, \langle\hat{n}(y,t)\rangle \\
\simeq& \Phi(x)\Phi^*(y) \langle \hat{\psi}^\dagger(x,t)\hat{\psi}(y,t)\rangle\\
&+\Phi(x)\Phi(y) \langle \hat{\psi}^\dagger(x,t)\hat{\psi}^\dagger(y,t)\rangle
+ \rm{c.c.}
\end{split}\end{equation}
In this equation, the symbol ``$:$'' denotes normal ordering and the
final expression is the Bogoliubov evaluation of $G_2$, encompassing
the effects of quantum fluctuations at leading order. At zero
temperature, the average $\langle\cdots\rangle$ in Eq. \eqref{eq3} is
taken over the state $|\rm{BH}\rangle$. Although this state is
thermodynamically unstable and cannot support a thermal distribution,
finite temperature effects can still be included as explained for
instance in Refs. \cite{Mac09,Rec09,Fab18}.

\begin{figure}
\includegraphics[width=\linewidth]{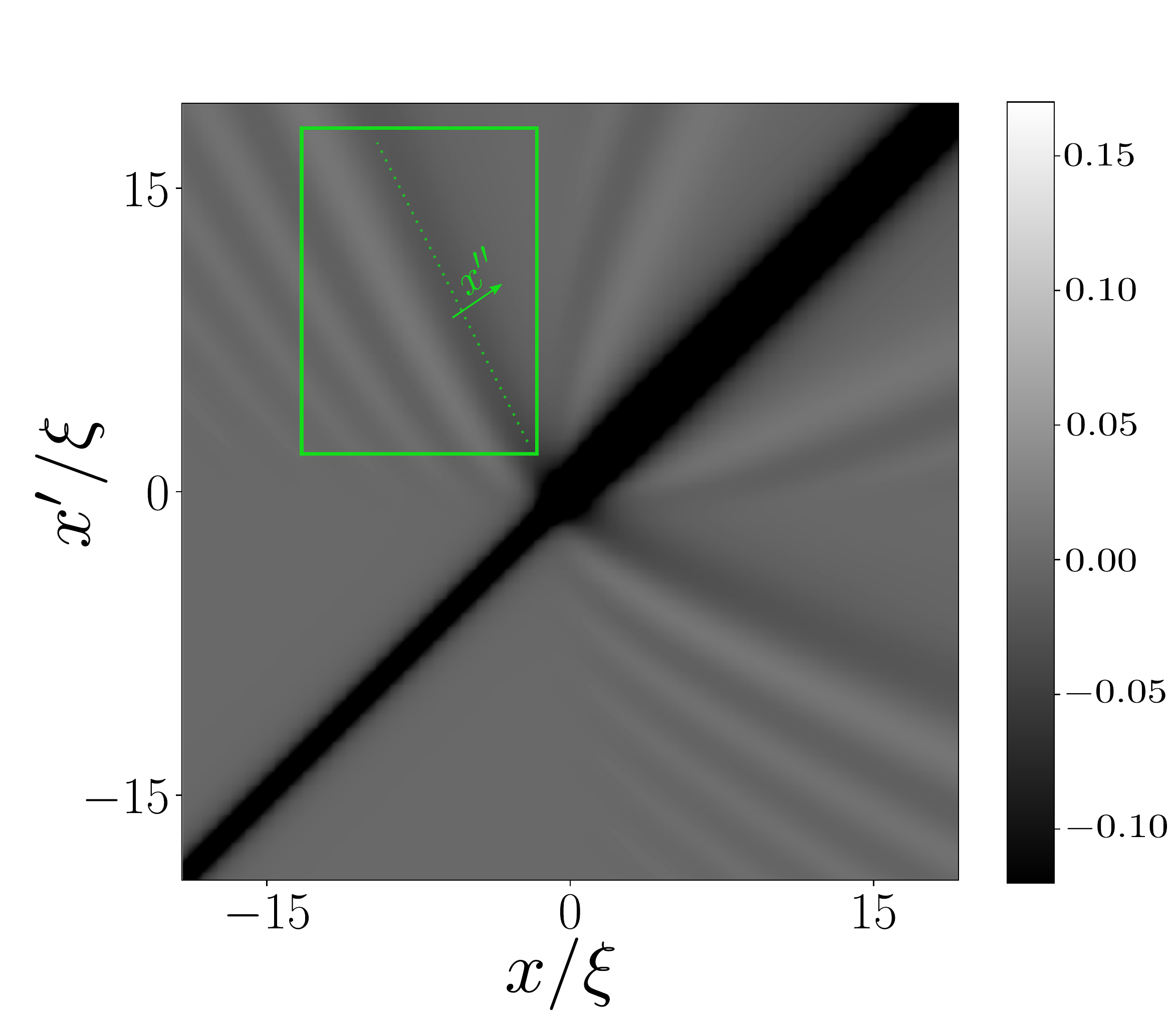}
\caption{Intensity plot of the dimensionless correlation function
$\xi\,( n_u n_d)^{-1/2} G_2(x,x')$ for
    $x$ and $x'$ close to the horizon. The parameter
    $\xi=\sqrt{\xi_u \, \xi_d}$ is the geometrical mean of the healing
    lengths $\xi_u$ and $\xi_d$, where $\xi_{(u/d)}=\hbar(m g n_{(u,d)})^{-1/2}$.
    The line of anti-correlation in the upper left and lower right
    quadrants corresponds to the merging close to the horizon of the
    Hawking-partner ($u|{\rm out}-d2|{\rm out}$) and Hawking-companion
    ($u|{\rm out}-d1|{\rm out}$) correlations. The green rectangle
    delimits the region where we average $G_2$ for
    comparison with experimental data (see Fig. \ref{fig3}).}
\label{fig2}
\end{figure}

In 2008 a collaboration between teams from Bologna and Trento
\cite{Bal08,Car08} pointed out that, in the presence of a horizon,
$G_2$ should exhibit non local features resulting from correlations
between the different outgoing channels, in particular between the
Hawking quantum and its partner ($u|{\rm out}-d2|{\rm out}$
correlation in our terminology). The importance of this remark lies in
the fact that, due to the weak Hawking temperature $T_{\sss\rm H}$ (at
best one fourth of the chemical potential \cite{Lar12}), the direct
Hawking radiation is expected to be hidden by thermal fluctuations,
whereas density correlations should survive temperature effects in
typical settings \cite{Rec09}. This idea has been used to analyze the
Hawking signal in Ref. \cite{Nov19}, where a stationary correlation
pattern was measured in the vicinity of the horizon. In this region,
it is important for a theoretical treatment to account for the
position-dependence of the background density and to include the
contribution of the evanescent channels in the expansion
\eqref{eq2}. We also checked that it is essential to take into account
the contribution of the zero modes to obtain a sensible global
description of the quantum fluctuations. The corresponding two
dimensional plot of the density correlation pattern is represented in
Fig. \ref{fig2}.
$G_2$ has been computed at zero temperature, for $V_d/c_d=2.90$, which
imposes $V_u/c_u=0.59$ \cite{Lar12,Supp}. This value is chosen to
reproduce the experimental configuration studied in
Ref. \cite{Nov19}. The dotted line in the upper left quadrant of
Fig. \ref{fig2} marks the anti-correlation curve which results from
the Hawking-partner ($u|{\rm out}-d2|{\rm out}$) and Hawking-companion
($u|{\rm out}-d1|{\rm out}$) correlations. We find that these two
correlation lines, which separate at large distance from the horizon
\cite{Car08,Rec09,Lar12}, merge close to the horizon, as
also observed experimentally.

\begin{figure}
\includegraphics[width=\linewidth]{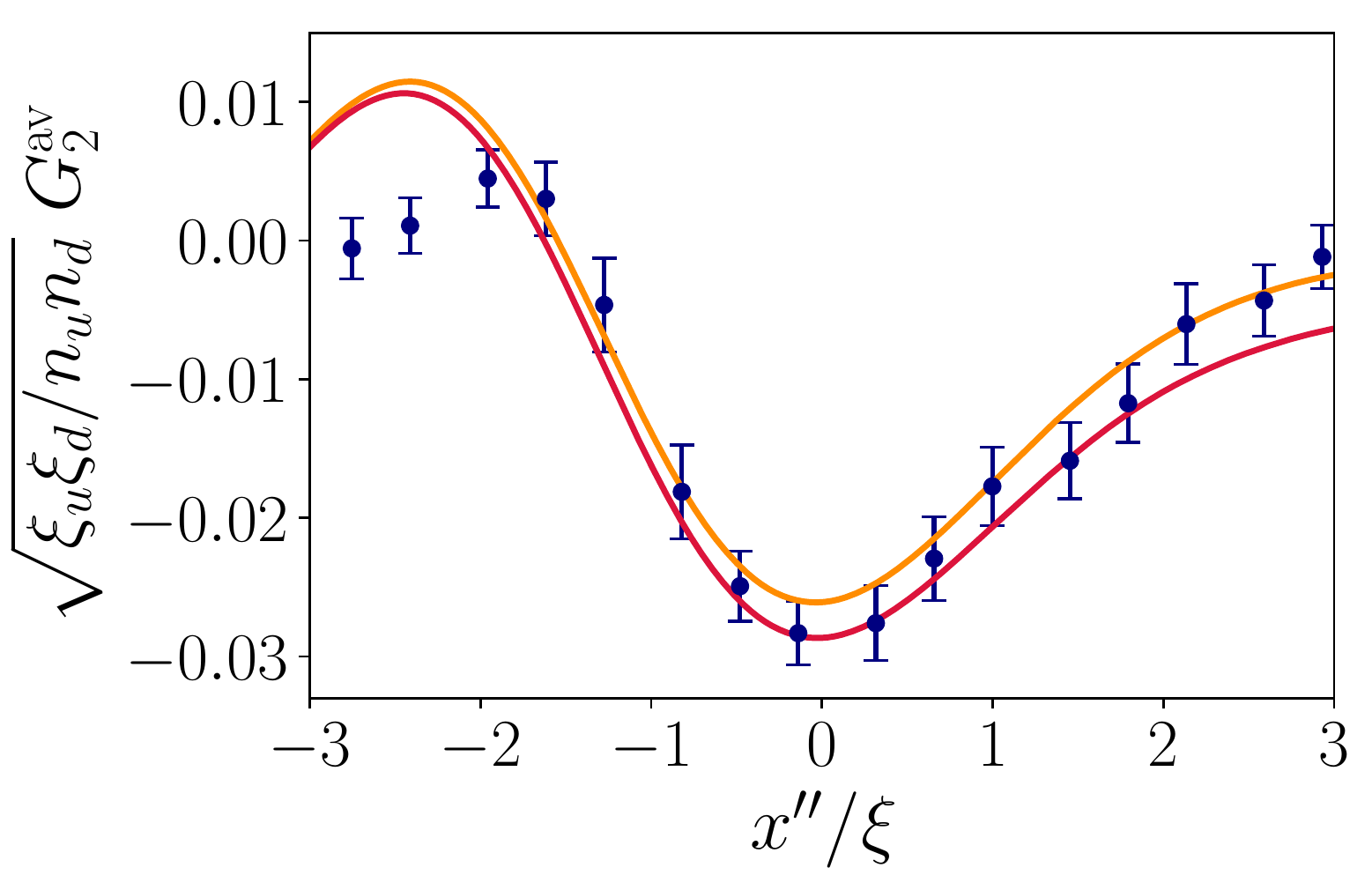}
\caption{Red solid line: zero temperature density correlation function
  ${G_2^{\rm av}}(x,x')$ plotted as a function of $x''$. The blue dots
  with error bars are the results of Ref. \cite{Nov19}. The orange
  solid line is the finite temperature result for $k_{\sss\rm
    B}T=0.2\, g n_u$, i.e., $T\simeq 1.9\, T_{\sss\rm H}$.}
\label{fig3}
\end{figure}

A precise comparison of our results with experiment can be achieved by
following the procedure used in Ref. \cite{Nov19}, which consists in
averaging $G_2$ over the region inside the green rectangle represented
in Fig. \ref{fig2}. One defines a local coordinate $x''$ which is
orthogonal to the locus of the minima of $G_2$, and one plots the
averaged $G_2$ (denoted as $G_2^{\rm av}$) as a function of the
variable $x''$. This is done in Fig. \ref{fig3}.
We insist that the good agreement between our approach and the
experimental results can only be achieved through a correct
description of the quantum fluctuations -- Eq. \eqref{eq2} --
including the contribution of zero modes and of evanescent channels.

It has been noticed by Steinhauer \cite{Ste15} that the determination
of $G_2(x,x')$ in the upper left (or lower right) 
quadrant of the $(x,x')$-plane makes
it possible to evaluate the Hawking temperature thanks to the
relation
\begin{equation}\label{eq4}
\begin{split}
& S_{u,d2}(\omega)S^*_{d2,d2}(\omega)=
\langle \hat{c}_{\sss U}(\omega) \hat{c}_{\sss D2}(\omega)\rangle =\\
& 
\frac{{\cal S }_0^{-1}}{ \sqrt{n_u n_d  L_u L_d}}
\int_{-L_u}^0 \!\!\!\!\! dx \!\int_0^{L_d} \!\!\! dx'
\, e^{-i(k_{\rm H} x+k_{\rm P} x')}G_2(x,x').
\end{split}
\end{equation}
In this expression $S$ is the matrix which describes the scattering of
the different channels onto each other, and ${\cal S }_0(\omega)
=(u_{k_{\rm
    H}}+v_{k_{\rm H}}) (u_{k_{\rm P}}+v_{k_{\rm P}})$ is the static
structure factor, where the $u_k$'s and the $v_k$'s are the standard
Bogoliubov amplitudes of excitations of momentum $k$ (see, e.g.,
Refs. \cite{PeSm,PiSt}). The $\hat{c}_{\sss L}$'s are outgoing modes
related to the incoming ones by the $S$-matrix \cite{Rec09}
\begin{equation}\label{eq5}
\begin{pmatrix}\hat{c}_{\sss U}\\ \hat{c}_{\sss D1}
\\ \hat{c}_{\sss D2}^\dagger\end{pmatrix}
=
\begin{pmatrix}
S_{u,u} & S_{u,d1} & S_{u,d2} \\
S_{d1,u} & S_{d1,d1} & S_{d1,d2} \\
S_{d2,u} & S_{d2,d1} & S_{d2,d2} \\
\end{pmatrix}
\begin{pmatrix}\hat{b}_{\sss U}\\ \hat{b}_{\sss D1}
\\ \hat{b}_{\sss D2}^\dagger\end{pmatrix}.
\end{equation}

\begin{figure}
\includegraphics[width=\linewidth]{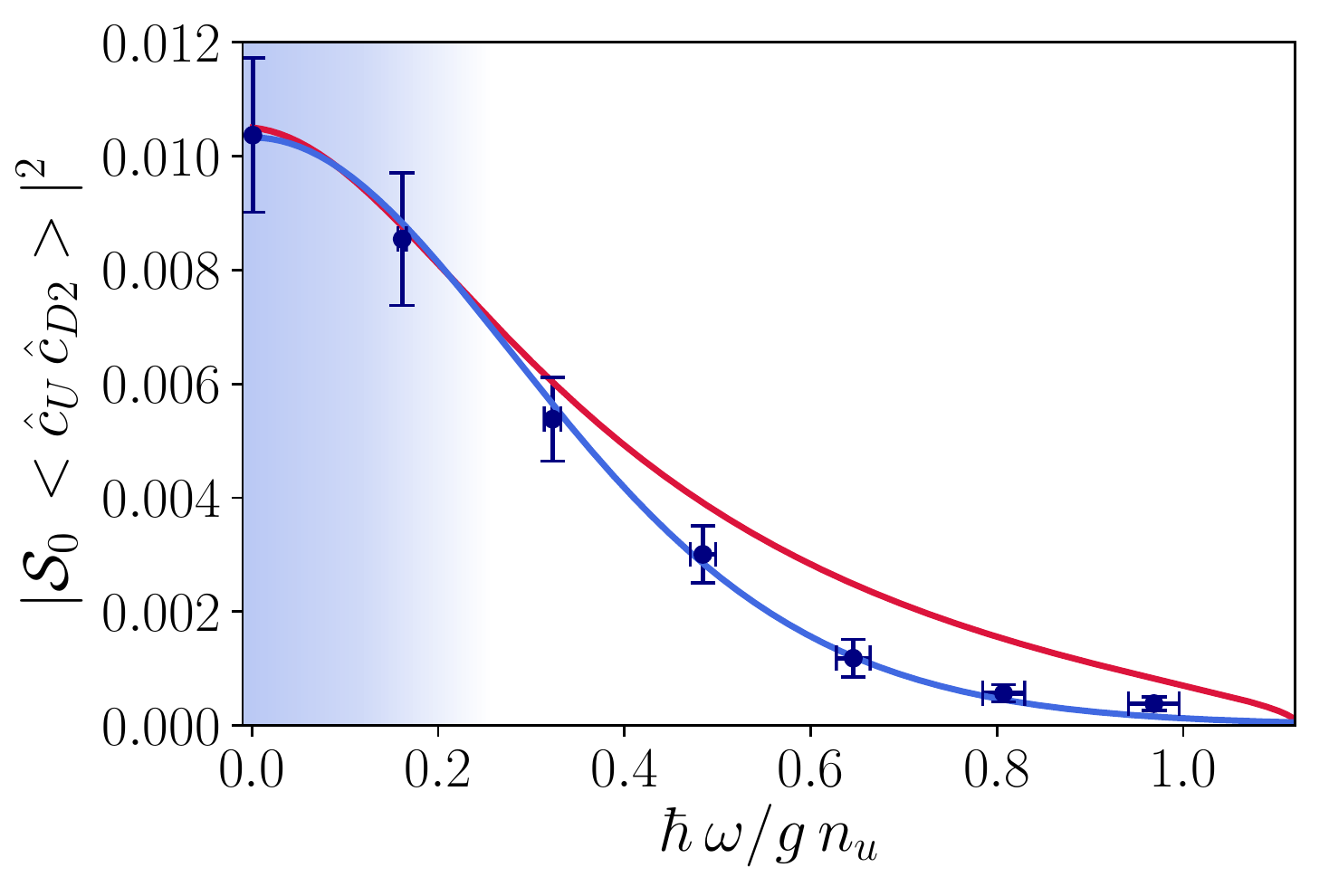}
\caption{Hawking-partner correlation signal represented as a function
  of the dimensionless energy. The red solid curve is the theoretical
  result from Eq. \eqref{eq4}. The dots with error bars are
  from Ref. \cite{Nov19}. They are obtained after processing the
  experimental result for $G_2$ by means of the Fourier transform
  \eqref{eq4}. The blue region corresponds to a domain where the
  ratio of Hawking and partner wavevectors is equal to its long
  wavelength value within a 10 \% accuracy. The blue solid curve is
  the theoretical result obtained by neglecting dispersive effects in
  Eq. \eqref{eq4} and discarding the contribution of the companion
  $d1|{\rm out}$ channel (see the text).}
\label{fig4}
\end{figure}

The Fourier transform of $G_2$ in Eq. \eqref{eq4} is performed at
fixed $\omega$, for wavevectors $k_{\rm H}(\omega)$ and $k_{\rm
  P}(\omega)$ which are the momenta relative to the condensate of a
Hawking quantum and its partner ($u|{\rm out}$ and $d2|{\rm out}$
channels in our terminology) having an energy $\hbar\omega$ in the
laboratory frame. The integration region $[-L_u,0]\times[0,L_d]$ lies
in the upper left quadrant of Fig. \ref{fig2}, and should be adapted
for each value of $\omega$ in such a way that \cite{Nov15,Fab18}
\begin{equation}\label{window}
\frac{L_u}{|V_{g,{\rm H}}(\omega)|} = \frac{L_d}{V_{g,{\rm P}}(\omega)},
\end{equation}
where $V_{g,{\rm H}}(\omega)$ [$V_{g,{\rm P}}(\omega)$] is the group
velocity of a Hawking quantum [of a partner] of energy
$\hbar\omega$. We have checked that once the prescription
\eqref{window} is fulfilled, formula \eqref{eq4} is very well verified
\cite{Supp}.  It is then intriguing to observe that, while theory and
experiment both agree on the value of $G_2$ in real space
(Fig. \ref{fig3}), they do not for the correlation $\langle
\hat{c}_{\sss U}(\omega) \hat{c}_{\sss D2}(\omega)\rangle$: as can bee
seen in Fig. \ref{fig4}, the agreement is restricted to the low energy
regime. This is the bluish region in the figure, which corresponds to
a domain where the ratio $k_{\rm H}(\omega)/k_{\rm P}(\omega)$ is
equal to its long wavelength value $(c_u-V_u)/(c_d-V_d)$ with an error
less that 10 \%.

Let us discuss this discrepancy in some detail. The interest of
Eq. \eqref{eq4} lies in the fact that the scattering matrix
coefficient $S_{u,d2}$ is the equivalent of the Hawking $\beta$
parameter: its squared modulus is expected to behave as a Bose thermal
distribution $n_{\sss T_{\rm H}}(\omega)$ with an effective
temperature $T_{\sss\rm H}$, the Hawking temperature \cite{Haw74}. In
an analogous system such as ours, because of dispersive effects, this
equivalence is only valid in the long wavelength limit, typically in
the blue region of Fig. \ref{fig4}.
This suggests a possible manner to reconcile theory and experiment: 
we assume that the ratio $k_{\rm H}(\omega)/k_{\rm P}(\omega)$ is
$\omega$-independent and equal to its low energy value,
$(c_u-V_u)/(c_d-V_d)$ (this value is denoted as $\tan\theta$ in
Refs. \cite{Ste16,Nov19}). We also assume that, in the scattering
process schematically illustrated in Fig. \ref{fig1} for the
$D2$-mode, the companion $d1|{\rm out}$ channel plays a negligible
role, so that the $|S_{d1,d2}|^2$ term can be omitted in the
normalization condition $|S_{d2,d2}|^2=1+|S_{u,d2}|^2+|S_{d1,d2}|^2$
of the $S$-matrix (see, e.g., Ref. \cite{Rec09}). Then one obtains
\begin{equation}\label{eq6}
|S_{u,d2}|^2 |S_{d2,d2}|^2 \simeq
n_{\sss T_{\rm H}}(\omega)[1+n_{\sss T_{\rm H}}(\omega)].
\end{equation}
Using the experimental values from Ref. \cite{Nov19} for $V_{\alpha}$
and $c_{\alpha}$ ($\alpha \in \{u, \, d\}$) and for the Hawking
temperature $T_{\sss\rm H}$ leads, within approximation \eqref{eq6},
to the blue curve of Fig. \ref{fig4} which agrees with the results
published in Ref. \cite{Nov19} (blue dots with error bars). It is
important to note that this procedure is self-consistent in the
following sense: If one performs numerically the Fourier transform
\eqref{eq4} over a domain which, instead of fulfilling the relation
\eqref{window}, verifies the $\omega$-independent condition
$L_u/|V_u-c_u|=L_d/(V_d-c_d)$ --appropriate in a non-dispersive, long
wavelength approximation-- one obtains a result (not shown for
legibility, but see \cite{Supp}) close to a thermal spectrum, i.e., to
the blue curve in Fig. \ref{fig4}.  Although this procedure is
self-consistent, it is not fully correct, as can be checked by the
fact that the resulting value of
$\langle \hat{c}_{\sss U}(\omega) \hat{c}_{\sss D2}(\omega)\rangle$
only agrees with the exact one (red curve in Fig. \ref{fig4}) in the
long wavelength limit. Stated differently: this procedure leads to the
erroneous conclusion that the radiation is fully thermal. However,
since all approaches coincide in the long wavelength regime (blue
colored region of Fig. \ref{fig4}), they all lead to the correct
determination of the Hawking temperature. For a flow with
$V_d/c_d=2.9$, our theoretical treatment yields
$k_{\sss\rm B}T_{\sss\rm H}/(g n_u)=0.106$, whereas the experimental
value reported for this quantity in Ref. \cite{Nov19} is 0.124
(corresponding to a Hawking temperature $T_{\sss\rm H} =0.35$ nK).

In conclusion, our work sheds a new light on the study of quantum
correlations around an analogous black hole horizon, and on the
corresponding Hawking temperature. From a theoretical point of view,
we argue that the contribution of zero modes is essential for
constructing a complete basis set necessary to obtain an accurate
description of the quantum fluctuations. This claim is supported by
the excellent agreement we obtain when comparing our results with
recent experimental ones. On the experimental side, we substantiate
the determination of the Hawking temperature presented in
Ref. \cite{Nov19}, although we find that the Hawking spectrum is not
thermal for all wavelengths. We identify a natural but unfounded
procedure for analyzing the information encoded in $G_2(x,x')$ which
leads to the opposite conclusion; we show that, within our approach,
an alternative analysis of the correlation pattern accurately accounts
for non-hydrodynamical effects. It would thus be interesting to re-analyze
the data published in Ref. \cite{Nov19} to investigate if the
windowing \eqref{window} we propose for Eq. \eqref{eq4} modifies the
experimental conclusion for the Hawking-partner correlation signal and
confirms the departure from thermality we predict.

\begin{acknowledgements}
We acknowledge fruitful discussions with I. Carusotto, M. Lewenstein,
and J. Steinhauer, whom we also thank for providing us with
his experimental data.
\end{acknowledgements}

\end{document}